# Guidestar-free image-guided wavefront-shaping

Tomer Yeminy,[1] and Ori Katz[1,*]

[1] *Department of Applied Physics, The Hebrew University of Jerusalem, Jerusalem, 9190401, Israel*
*\* orik@mail.huji.ac.il*

**Abstract:** Optical imaging through scattering media is a fundamental challenge in many applications. Recently, substantial breakthroughs such as imaging through biological tissues and looking around corners have been obtained by the use of wavefront-shaping approaches. However, these require an implanted guide-star for determining the wavefront correction, controlled coherent illumination, and most often raster scanning of the shaped focus. Alternative novel computational approaches that exploit speckle correlations, avoid guide-stars and wavefront control but are limited to small two-dimensional objects contained within the 'memory-effect' correlations range. Here, we present a new concept, *image-guided wavefront-shaping*, allowing non-invasive, guidestar-free, widefield, incoherent imaging through highly scattering layers, without illumination control. Most importantly, the wavefront-correction is found even for objects that are larger than the memory-effect range, by blindly optimizing image-quality metrics. We demonstrate imaging of extended objects through highly-scattering layers and multi-core fibers, paving the way for non-invasive imaging in various applications, from microscopy to endoscopy.

**Introduction**

High-resolution imaging and light control through highly scattering, turbid, media are fundamental problems that are important for a variety of applications, ranging from microscopy [1], through manipulation of cells and molecules [2], to astronomy [3]. When light propagates in a turbid medium, its wavefront is strongly distorted due to the spatially-varying refractive index, prohibiting the use of conventional imaging techniques. Great research efforts have led to the development of several approaches aimed at mitigating the effects of sample inhomogeneity. However, most of these approaches have limited success in highly-scattering turbid media: techniques that rely on unscattered, 'ballistic' photons, such as optical coherent tomography (OCT), confocal- and two-photon microscopy, are inefficient in thick highly-scattering media due to the exponential decay of ballistic photons at depths [1]. Adaptive-optics techniques [4], which aim at correcting low-order aberrations by the use of active elements such as deformable mirrors, are ineffective in highly-scattering media, since the number of scattered modes is orders of magnitude larger than the number of controllable degrees of freedom (DOFs) of the correction device. As a result, imaging and focusing light with these approaches are limited to depths of only a few hundred micrometers in soft tissue.

In the last decade, a paradigm shift in imaging and focusing light inside and through turbid media has been made following the introduction of 'wavefront-shaping' [5]–[15]. In wavefront-shaping, the illuminating or detected wavefront is shaped by a high-resolution spatial light modulator (SLM) using thousands of controlled DOFs to compensate for multiple-scattering induced distortions. In contrast to adaptive-optics, the wavefront-shaping correction cannot perfectly correct the scattering wavefront distortions, since the number of scattered modes exceeds by orders of magnitude even the large number of controlled DOFs of an SLM. However, the SLM-based correction still yields a diffraction-limited, high-contrast focus [9], which has allowed breakthroughs such as looking around corners [12], microscopic imaging [5], [7], [8], [16], and focusing light through tissues in space [7], and time [11].

While in the early demonstrations of wavefront-shaping the wavefront correction was found by direct invasive access to the object plane, located behind or inside the scattering medium [9], [12], in most practical applications such as biomedical imaging, such invasive procedures are impossible. Thus, the major challenge in the field today is in finding approaches that will allow to determine the wavefront-correction in a non-invasive, robust, and simple fashion, using only measurements performed *outside* of the scattering medium, and ideally without requiring labeling of the sample.

Despite the great recent advancements, the state-of-the-art noninvasive approaches for wavefront-shaping are severely restricted by requiring a guide-star to generate the feedback mechanism for the wavefront correction [7]. The large variety of developed guide-stars include the use of optical nonlinearities [16]–[18], which require external, and usually invasive labeling. Ultrasound-based guide-stars that rely on photo-acoustic [7], [19] or acousto-optic [13], [20], [21] feedback, come at the price of significantly reduced resolution compared to the optical diffraction-limit, and increase the system complexity. Very recently, several all-optical guide-star free approaches for focusing using wavefront-shaping have been presented. These approaches optimize a metric that relies either on the memory-effect [22] or on fluorescence variance optimization [23], [24] for focusing. However, these require controlled coherent illumination, fluorescence labelling and raster-scanning for imaging [23], [24], rendering these approaches inapplicable for imaging non-fluorescent samples, and necessitating acquisition times that are longer than the wavefront-correction time.

Alternatively, novel computational approaches that rely on the memory-effect for speckle correlations [25] avoid guide stars and wavefront-shaping altogether [26], [27]. These techniques computationally reconstruct the hidden objects using scattered light measurements [26], [27]. However, these methods work only for small isolated objects that are contained within the memory-effect's narrow angular range (the isoplanatic patch size), severely restricting their application in realistic scenarios where extended objects are usually concerned. Moreover, these computational approaches suffer from limited reconstruction fidelity and low probability of convergence when imaging complex objects [27].

Here, we present a guidestar-free, all-optical, noninvasive, and widefield, wavefront-shaping concept for direct imaging of hidden objects through highly scattering media, without any control on the illumination, labeling, or computational reconstruction. We avoid the strict usual requirement for a guide-star or known reference target [28] by adapting generalized image quality metrics as the feedback mechanism for determining the wavefront correction, effectively using the object as its own guide-star. Importantly, we show that image-based metrics can be used even when the initial scattered-light image is a low-contrast, seemingly featureless diffusive-halo of multiply scattered light (Fig.1b). Most importantly, our approach converges to the correcting wavefront even when the object extends beyond the memory-effect range, correctly imaging the object parts that are within the memory-effect angular range, a feat that is beyond the capabilities of correlations-based computational approaches [25], [29].

Our method is inspired by image-guided adaptive-optics [4], [30]–[32], where an image metric is used to find the wavefront correction for aberrations that are composed of several low-order modes. However, while image-guided adaptive-optics is effective in coping only with low order aberrations, the situation is profoundly different when combatting multiple scattering in turbid samples. Specifically, while low-order aberrations only mildly reduce the sharpness of the imaged objects, multiple scattering results in extremely low-contrast, seemingly-information-less diffusive blurs (Fig.1b), and conventional sharpness metrics may seem inappropriate. Another fundamental difference is that while aberrations are described by only a few tens to at most hundreds of parameters, in diffuse multiple-scattering the light is scattered to $\gg 10^6$ modes, and the wavefront-shaping correction requires the determination of thousands to hundreds of thousands of DOFs. As a result, the conventional starting point in adaptive-optics is a slightly blurred image and the final result is a nearly perfect image, whereas in wavefront-shaping the starting point is a seemingly information-less low-contrast diffusive light pattern, and the optimal end result is a sharp image over a speckled background (Fig.1b-c).

Nonetheless, we show that an image-based quality metric can robustly guide a wavefront-shaping correction with ~$10^4$ DOFs to directly image extended objects through highly scattering layers when the initial image is a low contrast diffusive blur. Strikingly, such guide-star free guidance leads to imaging even when the object extends beyond the memory-effect range. Moreover, we demonstrate that this new concept is general and useful in other applications by experimentally demonstrating widefield lensless endoscopic imaging through a commercial multicore fiber-bundle, without invasive access to the distal fiber end [33] or nonlinear measurements [34].

**Results**

Our concept for image-guided wavefront-shaping is presented in Fig. 1a. The goal is to image a hidden, incoherently-illuminated, object that is located behind a strongly scattering medium. A high-resolution SLM is placed on the other side of the scattering medium in order to undo the scattering-induced wavefront distortions [35]. The wavefront-shaped light from the SLM is Fourier-transformed by a single lens onto a high-resolution camera. To maximize the wavefront-correction field of view (FoV), the SLM plane is optically conjugated to the scattering medium surface [12], [35] using a 4-f telescope (not shown, see Supplementary material Section S1). Without correction, the scattered-light image (Fig. 1b) is of extremely low contrast, and results in a low-valued image quality metric. However, iteratively optimizing the SLM phase pattern to maximize an appropriate image quality metric, yields a diffraction-limited, high-contrast image of the hidden object (Fig. 1c), in the same manner as would be obtained by using an implanted point-source guide-star [12], [36]. Similarly to a correction using a guide-star, the angular FoV of the planar SLM correction is limited by the memory-effect to $\theta_{FoV} \approx \lambda / (2\pi L)$, where $L$ is the thickness of the diffusive medium [12]. Nonetheless, as we show below, image-guided wavefront-shaping can robustly determine the wavefront correction without a guide-star, even when the hidden object extends beyond the memory-effect FoV.

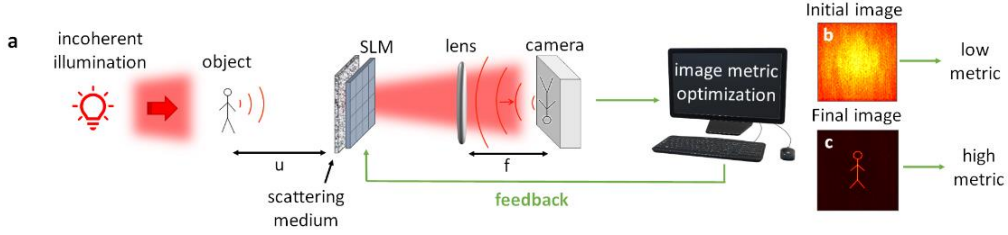

Fig. 1. Noninvasive imaging via image-guided wavefront-shaping, concept and numerical results: (a) An incoherently-illuminated object is hidden behind a highly-scattering medium. The scattered-light wavefront is shaped by a high-resolution SLM, which is conjugated to the scattering medium surface. The wavefront-shaped light is Fourier transformed by a lens on a high-resolution camera. (b) The initial camera-image is a low contrast diffusive blur. (c) Image-guided iterative optimization of the SLM phase correction aimed at maximizing a quality metric of the captured image yields a wavefront-correction with ~$10^4$ degrees of freedom, and a sharp, high-contrast widefield image of the hidden object.

To enable image-guided wavefront-shaping, two major challenges need to be resolved: 1) an image metric that can robustly lead from the initial image to the final image needs to be developed. This is especially challenging during the first steps of the optimization where the contrast and SNR are low; 2) An appropriate optimization algorithm that will ensure the convergence of ~$10^4$ DOFs (SLM phases) needs to be developed. Following in-depth numerical and experimental studies of image metrics, optimization algorithms, objects, and optical scattering samples, we show that these two challenges can be addressed by an adaptation of two well-established image quality metrics optimized by a genetic algorithm [37]. Specifically, we show that a combination of modified-entropy and variance metrics, along with a coarse-to-fine partitioning of the SLM during optimization, successfully copes with the low contrast and SNR of the diffused images in the first optimization steps, and converges to an appropriate wavefront-correction, yielding sharp high-contrast images through highly-scattering layers.

In order to understand and analyze the image formation in the system depicted in Fig.1a we consider the following mathematical model: within the memory-effect range [25], [29], a diffusive medium can be modeled as a thin random phase-screen, described by a spatial random phase $\phi(x,y)$ at a wavelength λ. Noting the SLM phase-correction by $\phi_{SLM}(x,y)$, the total accumulated phase of the corrected wavefront is $\phi(x,y)+\phi_{SLM}(x,y)$. For a point-object located on the optical axis at a distance $u$ from the scattering layer, the intensity distribution on the camera is:

$$PSF(x,y) \propto \left| F\left\{ \exp\left[ i\left( \frac{2\pi}{\lambda}\sqrt{u^2+x^2+y^2} + \phi(x,y) + \phi_{SLM}(x,y) \right) \right] \right\} \right|^2 \quad (1)$$

where $F\{\ \}$ stands for a 2D Fourier-transform performed optically by the lens (Fig.1a).

Without an SLM correction, i.e. for $\phi_{SLM}(x,y)=0$, the point-spread function, $PSF(x,y)$, is a random speckle pattern [27]. A perfect correction of both scattering and defocus terms is obtained for:

$$\phi_{SLM}^{opt}(x,y) = -\left[ \frac{2\pi}{\lambda}\sqrt{u^2+x^2+y^2} + \phi(x,y) \right] \quad (2)$$

For a point-object 'guide-star', this correction can be easily found by maximizing the intensity at a single camera pixel [12] or via phase-conjugation [36]. However, the case of an extended object cannot be solved in such a simple fashion (Fig. 2a-d). For the case of an incoherently-illuminated object that is contained within the memory-effect, the scattering PSF is isoplanatic and the resulting camera image is given by [27]:

$$I_{cam}(x,y) = I_{obj}(x,y) * PSF(x,y) \quad (3)$$

where $I_{obj}(x,y)$ is the intensity distribution at the object plane.

Since both $I_{obj}(x,y)$ and $PSF(x,y)$ are non-negative functions and the PSF without correction is a random wide-angle speckle pattern, $I_{cam}(x,y)$ before correction is a low-contrast diffusive blur (Fig. 1b). For a perfect SLM correction (Eq. 2) the PSF is a diffraction limited spot, which results in a high-contrast diffraction-limited image of the object (Fig. 1c). Hence, an image quality metric that takes into account sharpness and contrast can be used as a feedback for determining the SLM correction. The challenge is to find an object-independent and robust image metric to guide the SLM correction.

One may naively expect that maximizing a single pixel intensity [9], would lead to a perfect correction. However, a simple peak-intensity metric cannot distinguish between the desired correction that leads to a diffraction-limited PSF (Fig. 2b) and a correction that results in a 'matched-filter' PSF that is a mirror image of the object (Fig. 2c). This is a direct result of Eq. 2, since the intensity at a single pixel is the convolution of the object function with the PSF, i.e. the correlation of the object with $PSF(-x,-y)$. Maximizing the pixel intensity thus maximizes the correlation between the two functions, resulting in a PSF that is the mirror image of the object (see Supplementary information Section 5). This can be exemplified by considering the simple case of a two-point object, presented in Fig. 2a-d. For both the optimal correction with an intensity enhancement factor $\eta \equiv \max(PSF_{opt}(x,y))/mean(PSF_{init}(x,y))$ (Fig. 2b), and a matched-filter correction with enhancement factor $\eta/2$ (Fig. 2c) [9], the peak-intensity of the resulting image is the same (Fig. 2d), where $PSF_{opt}(x,y)$ and $PSF_{init}(x,y)$ are the initial and optimized PSFs, respectively.

In the past decades, numerous image metrics have been studied for image-guided adaptive-optics, with the earliest proposals by Muller and Buffington [30]. Two of the well-established metrics are entropy minimization and variance maximization [30]. Entropy minimization considers all image pixels rather than a single pixel as in peak optimization. Our in-depth numerical and experimental studies have shown that using a modified entropy, $H$, of the normalized camera image, leads to good imaging results for a large variety of objects and scattering layers (see Methods and Supplementary information Section 6). However, we found that this metric may sometimes converge to local minima, appearing as replications of the object in the vicinity of the true object's image (Fig. 1f). Unlike the modified-entropy, maximizing a variance metric (see Methods), optimizes the image contrast, and robustly converges to the object without any replications, under the condition that the region of interest (ROI) on the camera used for calculating the metric roughly matches the object dimensions. Otherwise, the large-area speckle background dominates the variance, hindering convergence to the object image.

Exploring various metrics, we have found that modified-entropy minimization followed by variance maximization over a small ROI that is extracted from the entropy optimization, results in sharp imaging through highly scattering media. The optimization process is explained and numerically demonstrated in Fig. 2e-g. The initial uncorrected image is presented in Fig. 1e. Modified-entropy minimization converges to a sharp image with nearby replications (Fig. 1f), from which the ROI for the variance optimization can be extracted. A following correction using variance maximization yields the desired corrected image (Fig. 1g).

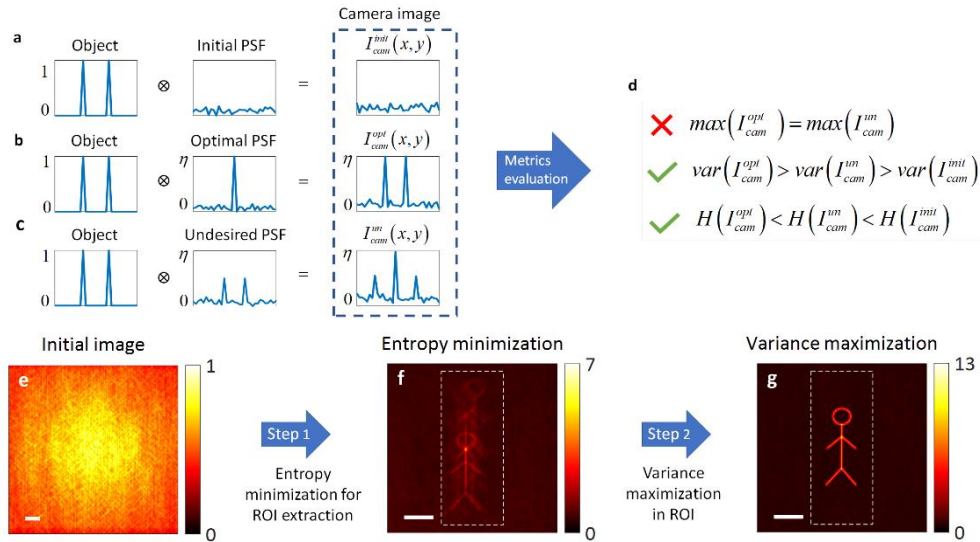

Fig. 2. Comparison of basic image quality metrics and description of the wavefront optimization process: (a-c) one-dimensional examples for imaging a simple two-point object using: (a) the initial uncorrected speckle PSF, (b) the desired wavefront-corrected diffraction-limited PSF, (c) an undesired 'matched-filter' PSF that resembles the object structure yields the same peak-intensity in the camera image as the desired diffraction-limited PSF, but does not recover the object. A desired image quality metric must distinguish between these two PSFs. (d) Comparison of three basic metrics for the simulated cases of (a-c): peak intensity, variance, and entropy (see Supplementary information Section 4). The variance and entropy successfully distinguish the desired PSF. (e-g) The two-step wavefront optimization process: (e) initial uncorrected camera image for a flat-phase SLM. (f) As a first step, the SLM wavefront correction is optimized to minimize the entropy of the camera image. This results in partial correction with a few replications of the object, from which a small region-of-interest (ROI, dashed line) for the second optimization step is extracted. (f) Variance maximization on the extracted ROI results in wavefront-correction and high-resolution imaging (g).

**Experimental imaging through scattering layers**

As an experimental proof-of-concept we imaged various incoherently-illuminated objects through highly-scattering optical diffusers. The results of two samples from these experiments using a single diffuser as the scattering medium are displayed in Fig. 3. These results demonstrate guide-star free, high-ontrast, widefield imaging of the hidden unknown objects, starting from initial very low-contrast scattered-light images.

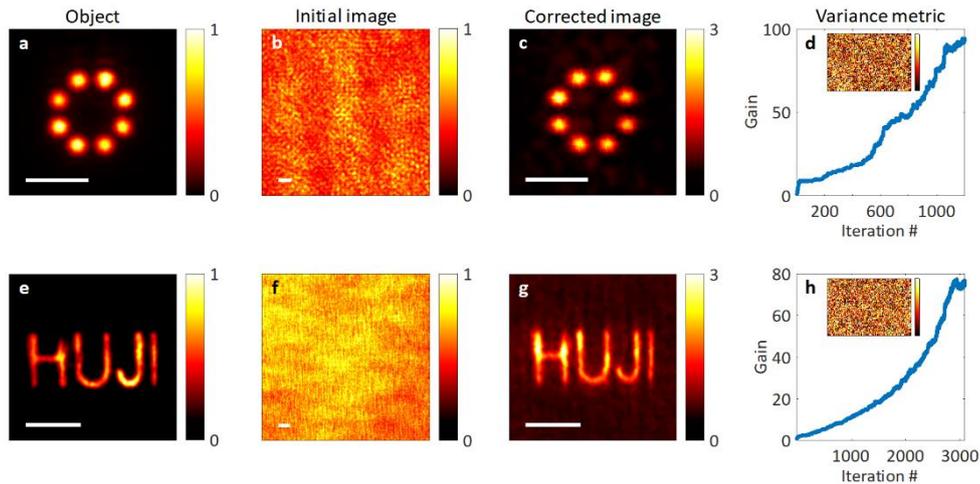

Fig. 3. Experimental imaging through a single highly scattering layer. (a) The object as imaged directly without the scattering layer. (b) The initial image through a highly scattering thin diffuser (60° scattering angle). (c) Camera image after image-guided wavefront optimization. (d) Evolution of the variance metric gain as a function of the optimization iterations. (e)-(h) Same as (a)-(d) for a different object and 20° diffuser. Scale bars: (a)-(c) 2mm, (e)-(g) 1.5mm. See Supplementary Video 1 for the evolution of the corrected image.

Guide-star free wide-field imaging through thin scattering layers, similar to those of Fig. 3, can also be obtained using computational approaches based on the memory-effect [26], [27]. However, these approaches fail when the object extends beyond the memory-effect, as is most often the case in practice. In contrast, image-guided wavefront-shaping does not suffer from this severe limitation, and robustly finds the wavefront correction for a single isoplanatic-patch even in such cases. This important and unique aspect of our approach is demonstrated numerically and experimentally in Fig. 4a-d and Fig. 4e-h, respectively. Fig. 4a-d presents a simulated scenario where the top and bottom halves of the scene (Fig. 4a) are scattered by uncorrelated scattering PSFs, each representing a single isoplanatic patch. Nonetheless, image-guided optimization robustly finds the correction for one of the object halves. Due to the stochasticity of the scattering and optimization process, a different half of the object is corrected in different numerical runs, but the optimization always results in imaging one isoplanatic patch. Fig. 4e-h experimentally demonstrates this capability, by imaging the same object used in Fig. 3e but through a scattering sample composed of a stack of two highly-scattering diffusers with a small spacing, yielding a memory-effect FoV that is considerably narrower than the object size. As expected, the optimization results in high-contrast imaging of the object parts that are contained within the memory-effect (Fig. 4g), a result that cannot be currently obtained by computational approaches without prior information on the hidden object.

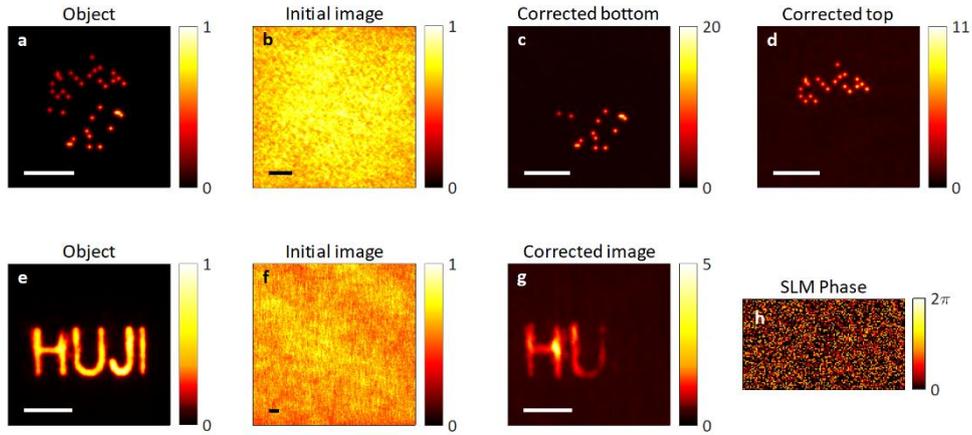

Fig. 4. Imaging extended objects through volumetric scattering with a limited memory-effect: (a-d) Numerical results for the case of an object whose top and bottom halves are scattered by uncorrelated scattering functions. (a) The hidden object. (b) Initial camera image with uncorrected wavefront. (c),(d) Final results of camera images from two independent optimizations, each randomly converges to a different wavefront correction matching a single isoplanatic patch (memory-effect field-of-view). (e-h) Experimental results through a scattering medium composed of two diffusers. (e) Object imaged without scattering. (f) Camera image without correction. (g) Camera image after image-guided optimization correct the left half of the object. (h) Final SLM phase pattern after optimization. Scale bars: (a)-(c) 1.1mm, (e)-(f) 1.5mm.

**Application in lensless endoscopy**

The concept of image-guided wavefront-shaping is general and can be applied not only to imaging through scattering layers, but also for e.g. lensless endoscopic imaging. Lensless endoscopes are a desired solution to minimally-invasive microendoscopy due to their reduced footprint and dynamic three-dimensional (3D) imaging [33], [38]. One attractive potential platform for lensless endoscopy is multicore fiber bundles, composed of thousands of individual cores [33], [34]. Conventionally, the fiber distal tip is placed adjacent to the imaged object, or conjugated to it using a distal lens. This results in a fixed focal plane, and a increased footprint. In order to image objects located away from the fiber distal facet without a distal lens, and in 3D, the random spatial phases of the different cores need to be corrected. Currently, this has only been performed by invasive access to the distal side [33] or by using nonlinear feedback signals [34]. However, the concept of image-guided wavefront-shaping and the analysis of Eqs.1-3 are valid also when the scattering medium is replaced by a fiber bundle. To experimentally demonstrate this capability, we replaced the scattering layer of Fig. 1a with a 50cm-long commercial multi-core fiber (Schott 191012-002), as detailed in Supplementary information Section 2. The results in Fig. 5b-e demonstrate the successful compensation of both the random intracore phase-distortions and defocus by image-guided wavefront shaping, leading to widefield imaging.

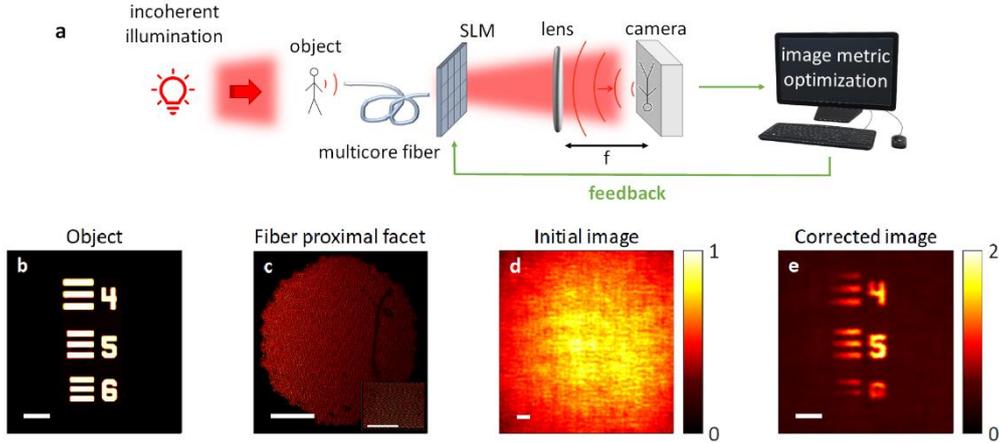

Fig. 5. Lensless widefield endoscopic imaging through a multi-core fiber bundle. (a) Schematic of the experiment: an incoherently illuminated object is placed at a small distance from the distal facet of a multicore fiber bundle having thousands of cores. The fiber proximal facet is conjugated to the SLM (imaging telescope omitted for simplicity) and Fourier-transformed on a camera. (b) The target object. (c) Conventional image of the fiber proximal facet does not reveal any object features. Inset: image of the cores conjugated to the SLM area. (d) The initial camera image without correction. (e) The final image after image-guided wavefront optimization, correcting both intracore phase distortion and defocus, resulting in a diffraction-limited image of the object. Scale bars: 0.25mm.

## Discussion

We have shown that guide-star free wavefront-shaping of extended objects through highly scattering layers can be obtained by image-guided wavefront-correction. For the planar objects considered in our proof-of-principle experiments, the correction does not only account for scattering distortions but also finds the correct axial focusing. This single correction can also be used for three-dimensional objects as it simultaneously corrects out-of-focus planes that are within the memory effect range [39].

As in other wavefront-shaping approaches, the intensity enhancement of the optimal correction is $\eta \approx \frac{\pi}{4} N_{SLM}$ [9], where $N_{SLM}$ is the number of controlled SLM pixels. The resulting optimized image contrast is approximately $\eta / N_{obj}$, where $N_{obj}$ is the number of bright resolution cells of the imaged object, assuming narrowband detection [12]. Thus, similar to other wavefront-shaping techniques, high-contrast images are obtained for mostly-dark objects or under darkfield imaging conditions. For broadband illumination or detection, the contrast of the initial and optimized images is proportional to $\delta\lambda / \Delta\lambda$, where $\Delta\lambda$ is the illumination bandwidth, and $\delta\lambda$ is the speckle spectral correlation bandwidth, assuming $\Delta\lambda > \delta\lambda$ [12].

In our proof of principle experiments we used liquid crystal SLMs, which resulted in rather long optimization times. Orders of magnitude faster optimization times can be reached by using faster SLMs, such as digital micromirror device (DMD). In addition, more efficient optimization algorithms can be employed to accelerate convergence.

Most importantly, state-of-the-art image metrics, taking into account prior or learned information on the imaged scene and the scattering medium [40] are expected to improve imaging of more complex objects, and potentially over larger FoVs, finding not only a single isoplanatic correction for each optimization (Fig. 4c-d), but possibly also simultanesouly correcting multiple isoplanatic patches. Merging wavefront shaping with the state-of-the-art computational approaches holds exciting potential for widefield imaging in various fields, ranging from biomedical to remote sensing applications.

## Methods

### Experimental setup

The complete experimental set-ups are detailed in Supplementary Figs. S1-S3. In these experiments the imaged objects were placed at distances of 1.5-40 cm behind the scattering sample and were illuminated by a spatially-incoherent narrowband pseudothermal source composed of a 20mW Helium-Neon laser (Thorlabs HNL210L) followed by a rotating diffuser. The SLM used was a Holoeye Pluto, and the scattered light was captured by a 5.5 megapixel Andor Zyla 5.5 scientific complementary metal–oxide–semiconductor camera, with bandpass filters for filtering the illumination wavelength (Thorlabs FB630-10, Semrock LL01-633-25 and FF01-650/SP-25).

The scattering medium in Fig. 2a-d and Fig. 2e-h was a diffuser with 60° and 20° scattering angle (Newport light shaping diffuser), respectively. In Fig. 3e-h the scattering medium was composed of a 10° diffuser placed 1mm away from another 10° diffuser (Newport light shaping diffuser). In Fig. 4d-g, we imaged a USAF test target (R1DS1N - Negative 1951) through a commercial Schott 191012-002, 50 cm long fiber bundle with 11k cores, pixel size of 7.5µm, and 0.83mm bundle diameter, and the target was placed <1.5cm from the fiber bundle. Camera exposure times were 70-400ms. All objects were transmission plates with different features.

**Image metrics and optimization**

The modified entropy metric, $H$, is calculated from the normalized camera image $\tilde{I}_{cam}[m,n] = \alpha I_{cam}[m,n] / \max(I_{cam}[m,n])$ by: $H[\tilde{I}_{cam}] = \sum_{m=1}^{M}\sum_{n=1}^{N} \tilde{I}_{cam}[m,n] \log(\tilde{I}_{cam}[m,n])$, where $I_{cam}[m,n]$ is $I_{cam}(x,y)$ sampled by the camera, and $\alpha = 0.4$, and the sum is performed over all camera pixels.

The variance metric is calculated from the camera image by:
$$V[I_{cam}] = \frac{1}{M_{ROI} \cdot N_{ROI}} \sum_{m=1}^{M_{ROI}}\sum_{n=1}^{N_{ROI}} I_{cam}^2[m,n] - \left(\frac{1}{M_{ROI} \cdot N_{ROI}} \sum_{m=1}^{M_{ROI}}\sum_{n=1}^{N_{ROI}} I_{cam}[m,n]\right)^2$$
, where the sum is performed over a small ROI that is extracted from the entropy optimization and fits the object image size.

The image-based sharpness metrics were optimized by a genetic algorithm [37] with a population size of 50 and 100 for Figs. 3, 5 and Figs. 1, 2, 4 respectively, initial and final mutation rates of $R_0 = 0.3$ and $R_{end} = 0.013$ respectively, and decay factor of $10^3$. We used coarse-to-fine partitioning of the SLM during the optimization process, gradually refining macro pixels from 120x120 to 15x15, where in each refinement step the macro pixel size in each dimension was divided by factor 2. A refinement step occurs when the metric changes in less than 5% over 50 generations.

**Simulations**

Numerical simulations were performed according to Equations 1-3. The scattering layers were modeled as a random phase plate with 256×256 random phases. The SLM was modeled as a 64×64 controlled phase modulator covering the entire random phase plate. For all simulations Gaussian noise was added to the raw camera images with a signal to noise ratio of 20 for Fig. 1d-g and 100 for Fig. 4a-d. A study of the effects of SNR can be found in Supplementary information Section 8.

**Funding**

This work is funded by the European Research Council (ERC) Horizon 2020 research and innovation program (grant no. 677909), Israeli Ministry of Science and Technology (Grant 712845), Human Frontiers Science Program (Grant RGP0015/2016).

**Disclosures**

The authors declare that there are no conflicts of interest related to this article.

**Acknowledgments**

We would like to thank Prof. Raanan Fattal and Prof. Yedid Hoshen from The Benin School of Computer Science and Engineering for fruitful discussions. In addition, we thank Avner She'altiely, Shlomo Gabay, and the staff from the Fine Mechanical Workshop of the Racah Institute of Physics.